# Measuring Metrics - A forty year longitudinal cross-validation of citations, downloads, and peer review in Astrophysics

**Michael J. Kurtz**
**Edwin A. Henneken**
Harvard-Smithsonian Center for Astrophysics

Citation measures, and newer altmetric measures such as downloads are now commonly used to inform personnel decisions. How well do or can these measures measure or predict the past, current of future scholarly performance of an individual? Using data from the Smithsonian/NASA Astrophysics Data System we analyze the publication, citation, download, and distinction histories of a cohort of 922 individuals who received a U.S. PhD in astronomy in the period 1972-1976. By examining the same and different measures at the same and different times for the same individuals we are able to show the capabilities and limitations of each measure. Because the distributions are lognormal measurement uncertainties are multiplicative; we show that in order to state with 95% confidence that one person's citations and/or downloads are significantly higher than another person's, the log difference in the ratio of counts must be at least 0.3dex, which corresponds to a multiplicative factor of two.

Introduction

Arguably one of the most important use of bibliometric techniques is in aiding in the evaluation of individual scholars. If the capability of individuals to contribute to an organization or discipline follows a Zipf (1949) or Lotka (1926) like power law distribution, it is obviously crucial for organizations to choose the most capable or talented for hiring, advancement and rewards. Simply applying the Pareto Principle, also known as the "80-20 rule", implies that a well-chosen faculty can be a factor of ten times as productive as a randomly chosen one.

At early career stages personnel decisions are essentially predictions about an individual's future achievements. Bibliometric methods, such as article and/or citation counts (Garfield, Sher and Torpie, 1964) have long been used to help evaluate individuals, and predict their future success; famously Garfield and Malin (1968) used citation counts to predict that Murray Gell-Mann would win the Nobel Prize in Physics, which he did the next year.

Recently improvements to raw citation counts, and non-citation (alternative or alt, e.g. Priem 2014) based measures have been suggested by many authors, article downloads (Kurtz and Bollen 2010; Haustein 2013) being the most common. How stable/accurate are these measures, and how well do these measures predict the future performance of individuals? In this paper we cross compare three substantially different indicators/predictors for an observational cohort of all 922 persons who received a PhD in Astronomy from a U.S. university in the five-year span 1972-1976.

The main three indicators in this study are the citation based Tori statistic (Pepe and Kurtz, 2012) which removes self-citations and normalizes citation counts by both the number of authors on a cited paper and the number of references in the citing paper; the downloads based Read10 statistic (Kurtz, et al 2005b: K05b) which counts current downloads of papers published within ten years of the download date, normalized by number of authors; and the combined results of elections and prize committee deliberations of the American Astronomical Society and the National Academy of Science.

In all of science there is no such thing as the one "true" indicator. All measures have error, and all measures are indirect. A common thermometer, for example, measures the length of a column of mercury, which must be calibrated and validated by comparison with other measurements. From the mercury column length one infers the temperature, which itself is just an approximate description of the ensemble of energy states of the particles which make up the object being measured; and which assumes that the distribution of energy states are in an ideal form (thermal equilibrium) which, in practice, is never exactly achieved. None the less temperature is a useful concept, and thermometers are useful instruments to measure it.

Measuring an individual's scholarly performance, potential, or ability is likewise an indirect process. Scholarly ability, itself, is a quality (Pirsig 1974) which can be perceived, but which is very difficult to specify exactly. The measurement of an individual's scholarly ability is often made by observing the accumulated actions of individual peer scholars. A peer scholar may vote to honor an individual, may choose to cite one of an individual's papers, and may choose to read one of an individual's papers.

This paper is not intended to compare different citation measures. The (near) full career retrospective data used here, of individuals from a single age cohort in a single discipline, are much less susceptible to the large variance in the per article number of authors and references than a modern, cross-disciplinary data set would be. The Tori was designed to address the more modern issue; for the data in this paper we expect the results using Tori to be very similar to any other citation based indicator, such as total citations (Garfield, Sher and Torpie 1964, co-author number normalized citations (K05b), h (Hirsch 2005), or the CWTS crown indicator (Moed, De Bruis and Van Leeuwen 1995; Moed 2010). Verification of this expection falls outside the scope of this paper.

There is a vast literature on citations and the comparison of various citation based measures, Moed (2005), Nicolaisen (2007), and Waltman (2015) offer reviews; Bornmann and Marx (2013) propose standards for using citations to measure individuals. Bornmann (2011) reviews peer review, and Agha and Li (2015) and Egghe and Bornmann (2013) compare the results of peer review with later citation information. Kurtz and Bollen (2010) and Moed and Halaki (2015) review article downloads, and discuss their relation to citations. Recently Wilsdon, et al (2015) have released a massive (463 pages) review on the role of metrics in research assessment; this included a correlation analysis of a large number of bibliometric and alt-metric indicators with each other, and with confidential peer evaluations from the U.K. Research Excellence Framework process over a very wide range of disciplines.

We perform a detailed examination on a well understood, complete sample, paying particular attention to the effects of selection, activity, age, (sub-)field, (co-)authorship, career path, and (crudely) publication type on the measurements. Additionally our download statistics are dominated by use by research astronomers, which removes another substantial complication. It has long been known that readership patterns match citation patterns only when the readers are active researchers in the same field (i.e. authors)  (Stankus and Rice 1982; see Kurtz and Bollen 2010 for a review).

Taking a complete, age defined sample of potential research astronomers we examine, at different stages in their careers, three different means by which peer groups of research astronomers indicate their respect for the research of an individual in the cohort.  Research astronomers decide to give an honor to a cohort member, research astronomers decide to cite an article by a cohort member, and research astronomers decide to read an article by a cohort member.

In the sections that follow we first describe the data we use; we show how bibliometric indicators are affected by age and activity sample bias;  we examine the ability of early career citations to predict mid career citations; we show that there is a log normal relation with a well defined mean and standard deviation; we develop a (near) bias free sample of career astronomy researchers, and show the pure relationships between pairs of indicators at the same and different times; we compare the ability of early citation measures to predict mid career download measures, and their ability to predict late career distinctions with the ability of early career distinctions to predict late career distinctions; we show that using a combination of measures we can discover a systematic bias in the election of astronomers to the U.S. National Academy of Science.  We conclude by discussing our results.

Whenever logarithms are used in this paper they are base 10.  Another convention is our use of the word "download": a download is considered as any accessing of data on a paper, whether full text, abstract, citations, references, associated data, or one of several other lesser-used options; this both increases the signal and removes much of any possible systematic due to access controls on the full text.

**Data**

We searched the Astronomy database of the Smithsonian/NASA Astrophysics Data System (ADS; Kurtz et al 2005a: K05a) for PhD Theses from U.S. Universities granted in the five years between 1972 and 1976.  We found 922, corresponding to 922 individuals.  B. Elwell performed this work in 2002.  Individuals in this cohort are now (2015) typically between 66 and 76 years old.

We used the ADS astronomy database to obtain a listing of all of each individual's astronomy papers.  We eliminated papers by persons with similar names, and added papers missed due to name changes (marriage) by hand curation.  This process required a substantial number of telephone calls.  Additionally we noted whether each individual was a member of the American Astronomical Society (AAS) according to the 2002 member directory.

These article lists do not include papers from the ADS Physics database; this very substantially reduced the scale of the homonym problem, at the cost of an underestimation of the impact of individuals whose career had a substantial physics component.  A detailed examination of several dozen individuals suggests that this has had no effect on any of our results.

We develop two measures using download statistics, Read10 and RQuot.  To compute the Read10 statistic we counted downloads of papers published in 1991 or later using download information from the ADS logs for the years 2000 and 2001.  At that time the ADS was heavily used by professional astronomers, worldwide (K05a) but had not yet been discovered by others, thus the ADS logs from then represent a reasonably fair and pure sample of what was being read by professional astronomers, without additional filtering. The RQuot measure is computed by subtracting Read10 from the total normalized downloads; this gives the reads in 2000-2001 of papers published before 1991. We correct this for the differing PhD dates of the cohort, using the K05b reads obsolescence; this correction is +/- 2% or smaller.

Additionally we developed Read10/14 and RQuot14 that use identical lists of papers as Read10 and RQuot, but use download information from the ADS logs for the years in 2012 and 2013. Because by 2012 the ADS was more widely used we filtered the logs by usage pattern to only include use by active scientists.

In 2000 the ADS covered planetary science less well than other types of astronomy; this creates a systematic problem with the reads data that will be addressed below.

In the year 2000 (1991) members of our PhD cohort were typically between 51(42) and 61(52) years of age.

The Read10 statistic used here is exactly the same as that used in K05b, which allows the plots in this paper to be directly compared with the plots in that paper. This definition differs in several details with that currently in use by the ADS (ADS 2014); the results of this paper are not affected by these differences.

For the citation analysis we use the ADS citation database (as of mid 2014); Abt (2006) has compared the completeness of the ADS citations with the Web of Science, both have a high level of completeness, but ADS is more complete in covering the astronomy conference literature and observatory publications, and WoS is (much) more complete in covering references originating outside physics and astronomy.

Because the citation data contains the publication date of the citing article, it is a simple matter to derive retrospectively the citation measure for any article at any date; this was done to compute the Tori statistic five (Tori5), twelve (Tori12), and twenty-five (Tori25) years past the PhD.

We also developed Tori10/14, which measures the citations in mid 2014 to those papers used to compute Read10 and Read10/14.

Membership in the American Astronomical Society (AAS) as of 1 Jan 2002 was taken from the 2002 AAS Membership Directory, 339 members of the cohort were in the AAS at that time; lists of AAS prize recipients and officers were taken from the AAS' web site (AAS 2014).

Membership in the U.S. National Academy of Science was taken from their website, along with the date elected, 14 members of the cohort have been elected into the NAS.

In addition to simple membership we have created three additional subsets of individuals. The category "honored" is used for persons who (as of 2014) have been elected to the NAS, or to any office in the AAS, or who have received any prize from the AAS, 29 from the cohort. Individuals in the category "highly honored" are people who have been elected to the NAS, or to the AAS presidency or vice presidency, or who have received a mid or late career prize from the AAS, 21 from the cohort. Finally, "young award winners" are individuals who have received one of the three prizes given by the AAS between 1970 and 1990 to individuals within five years of receiving the PhD, this comprises 49 individuals, most not in the cohort. More than half of the high honors were received after 2000.

**Sample Bias**

All bibliometric investigations are limited by the nature of the samples taken; this is especially true for comparisons between individual researchers. Some research disciplines have citation rates that differ from others by a factor of five (Schubert & Braun 1986; Radicchi, Fortunato and Castellano 2008) and citation measures for individuals increase approximately quadratically with the scientific age of the person. Thus comparing individuals or groups of individuals at different ages and from different fields is problematic. By choosing a sample of individuals from a single discipline and with very similar scientific ages we ameliorate these problems.

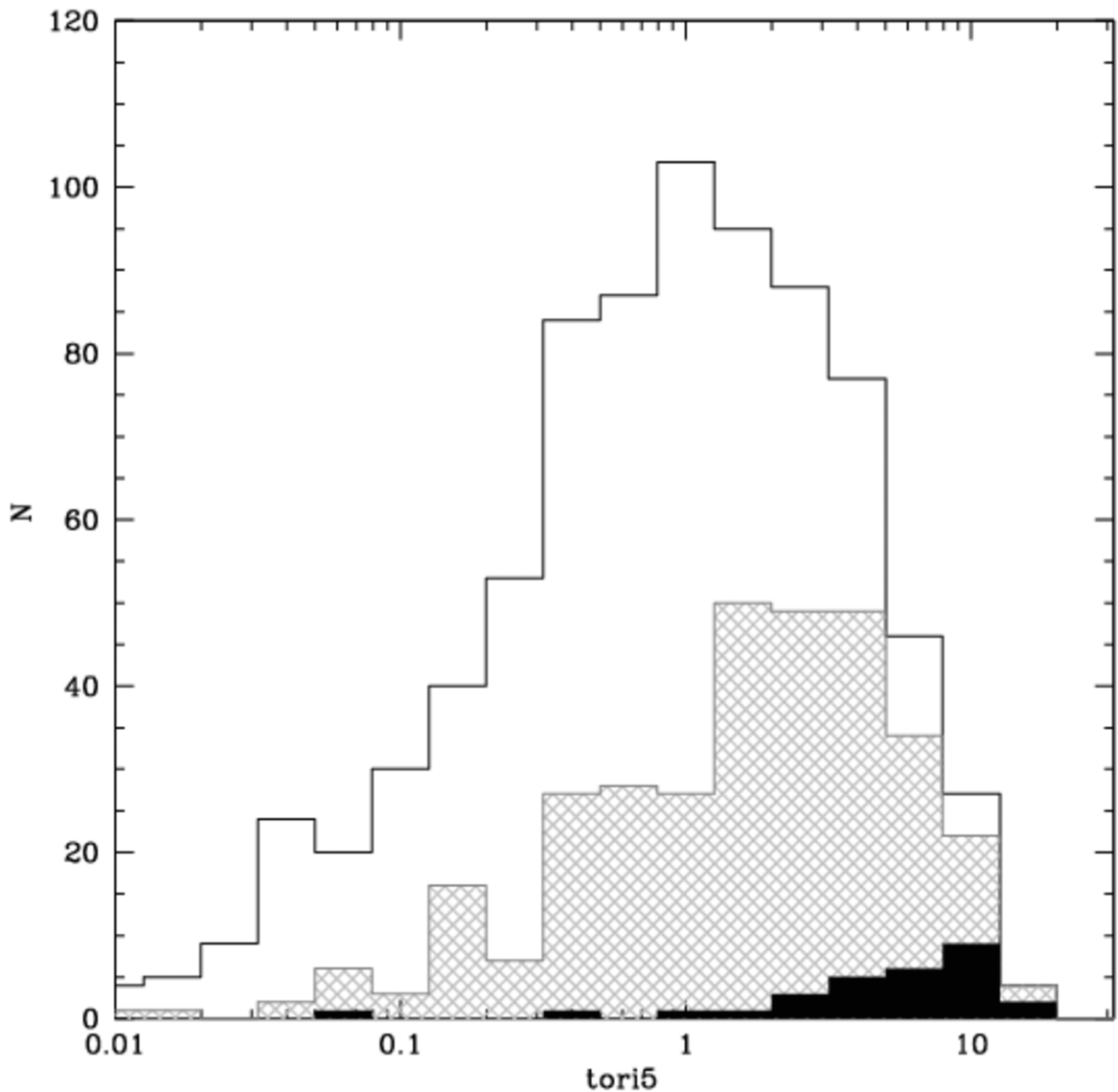

Figure 1. The effects of sample selection on the distribution of Tori5; clear: full sample, cross-hatched: 2002 AAS members, black: honored.

Additionally substantial difficulties can arise in determining the extent of the low productivity end of the distribution; this is sometimes known as survivor bias, which is a type of selection bias and refers to the logical error of inadvertently overlooking people with a lack of visibility. Most sample selection strategies systematically under sample low achievers. By taking all (U.S.) astronomy PhDs from our five-year cohort our sample approximates everyone at the relevant age that could be an author on an astronomy research paper.

Figures 1-3 show the distributions of Tori5, the citation based measure taken five years past the PhD, Tori25, the citation based measure taken 25 years past the PhD, and Read10, the measure of downloads of recently published articles (as of 2002), for the entire sample, and two subsets. The distributions are shown as histograms binned in equal logarithmic steps (0.2 dex, a multiplicative factor of 1.585). The outer, clear histogram shows the full sample, the crosshatched histogram shows the 2002 AAS member sample, and the black histogram shows the honored sample. One member of the honored sample was not a member of the AAS in 2002 (this person is currently an AAS member), otherwise these are nested proper subsets.

The distributions of Tori5 (figure 1) for the three samples are clearly different. At five years past the PhD individuals who would 22 (+/- 3) years later be members of the AAS (344 persons) tended to have higher citation counts than the group as a whole (922 persons) and substantially higher than those (578) who would not. Note that 125 persons (18 AAS members and 107 non-members) had Tori5 scores of zero; they had not received a citation five years past the PhD, and are not represented in the figure. The individuals who would later be honored are even more highly skewed toward high Tori5 scores than AAS members.

Tori25 (figure 2) is distributed similarly to Tori5, the concentration of the AAS members and of the honored subset is somewhat more pronounced. 36 individuals who had not been cited five years past the PhD were cited by 25 years past, leaving 89 persons (12 in the AAS sample, 77 not) who never received a citation

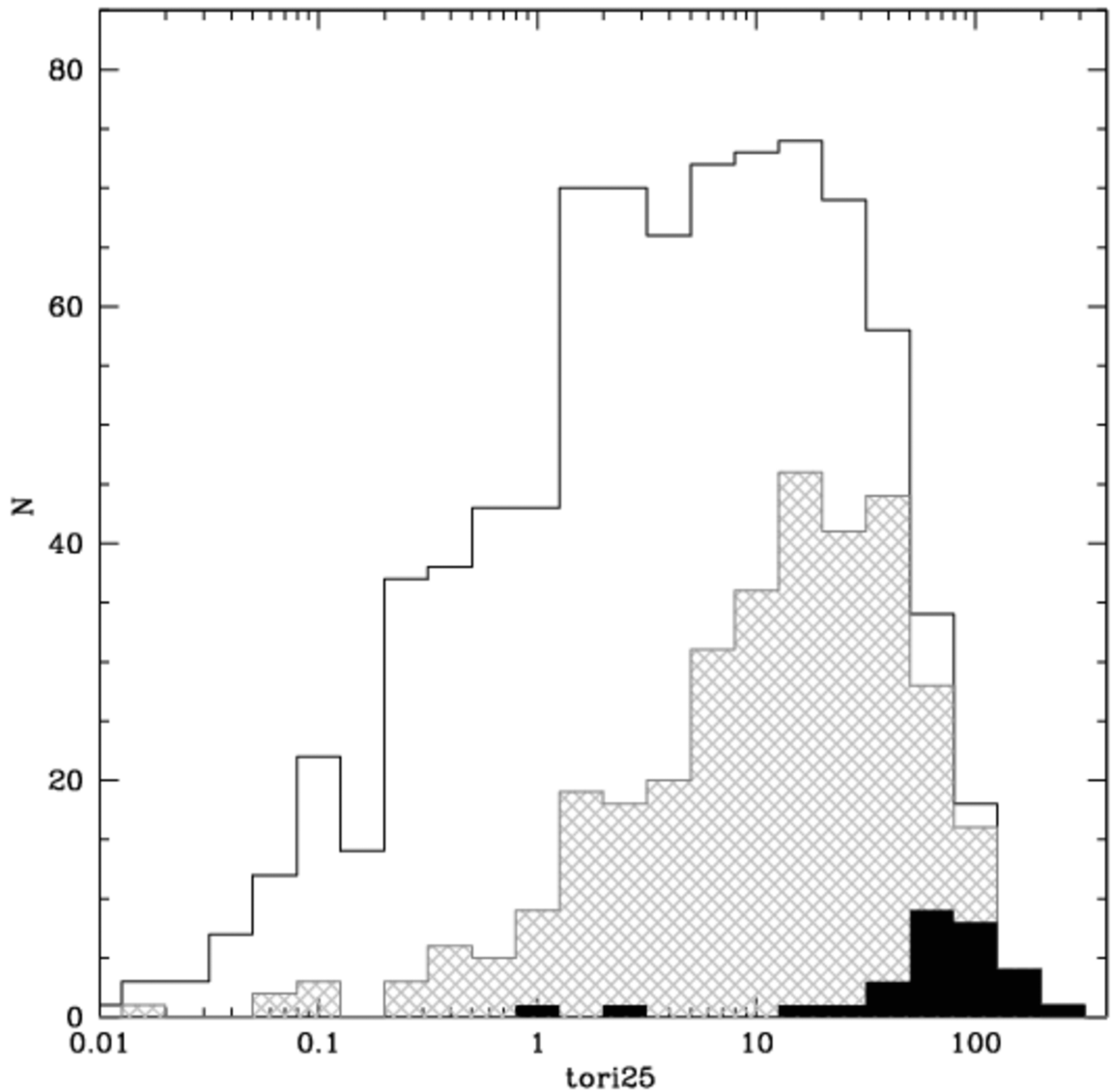

Figure 2. The effects of sample selection of the distribution of Tori25. Samples same as in figure 1.

The Read10 distribution is very different, 45% of the entire sample (418 total, 45 AAS members, 373 not) did not have a single download during 2000-2001 of a paper s/he published after 1990; they are not present in the diagram. The distribution of the remaining individuals is substantially more skewed toward higher production.

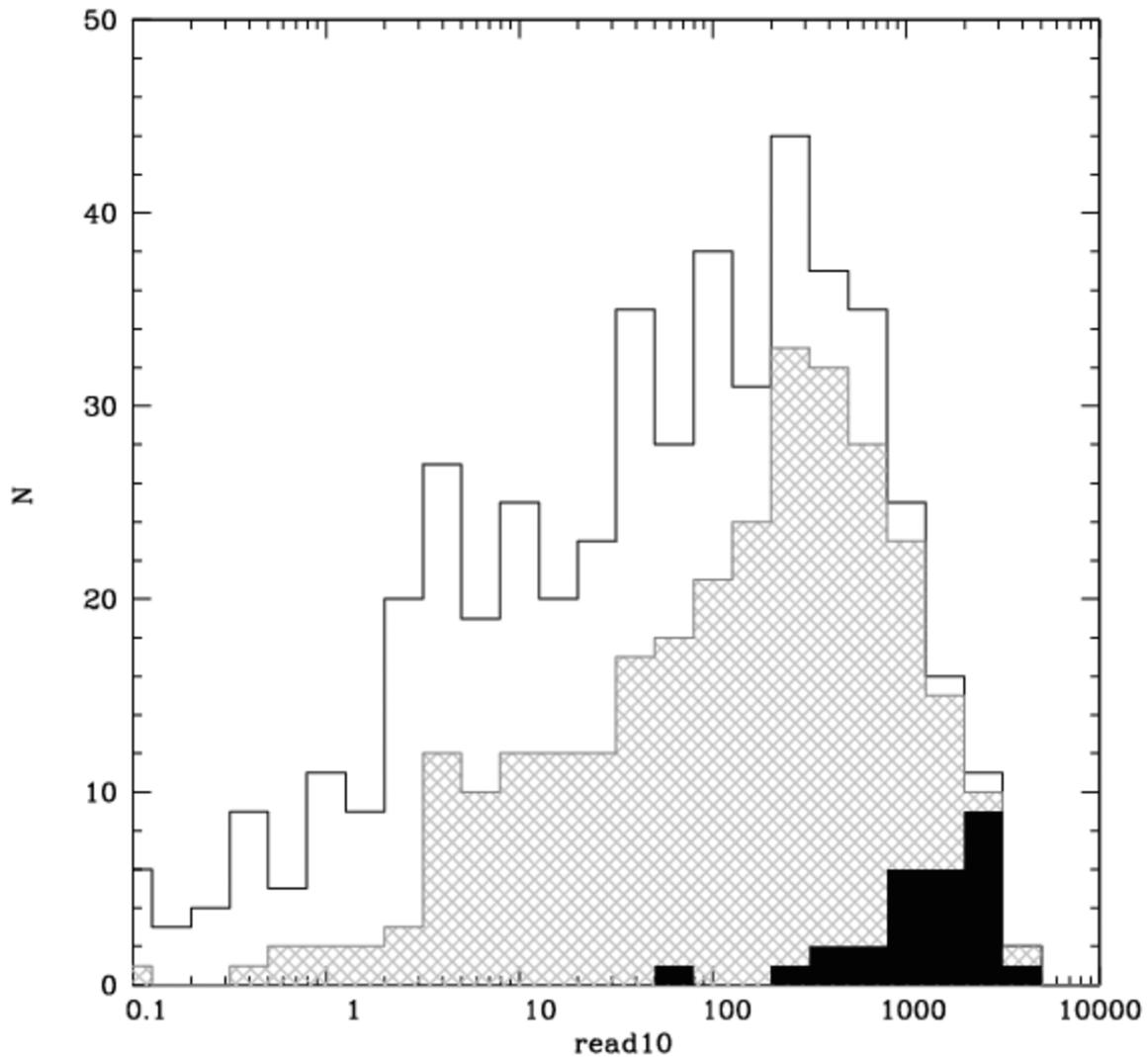

Figure 3. The effects of sample selection on the distribution of Read10. Samples same as in figure 1.

These three figures (1-3) taken together show the possible effects of different sampling strategies. If one does not take age into account the result is similar to merging figures 1 and 2; the ability to judge individuals from their position in such a merged distribution would be almost totally lost. Choosing a sample based on membership in a scholarly society, or a list of honored individuals, or some other measure of achievement results in results differing by the exact nature of the selection, as can be seen in all the diagrams. Choosing a sample from authors of recent publications (essentially the selection of figure 3) also induces a survivor bias, quite similar to that which membership in a scholarly society induces.

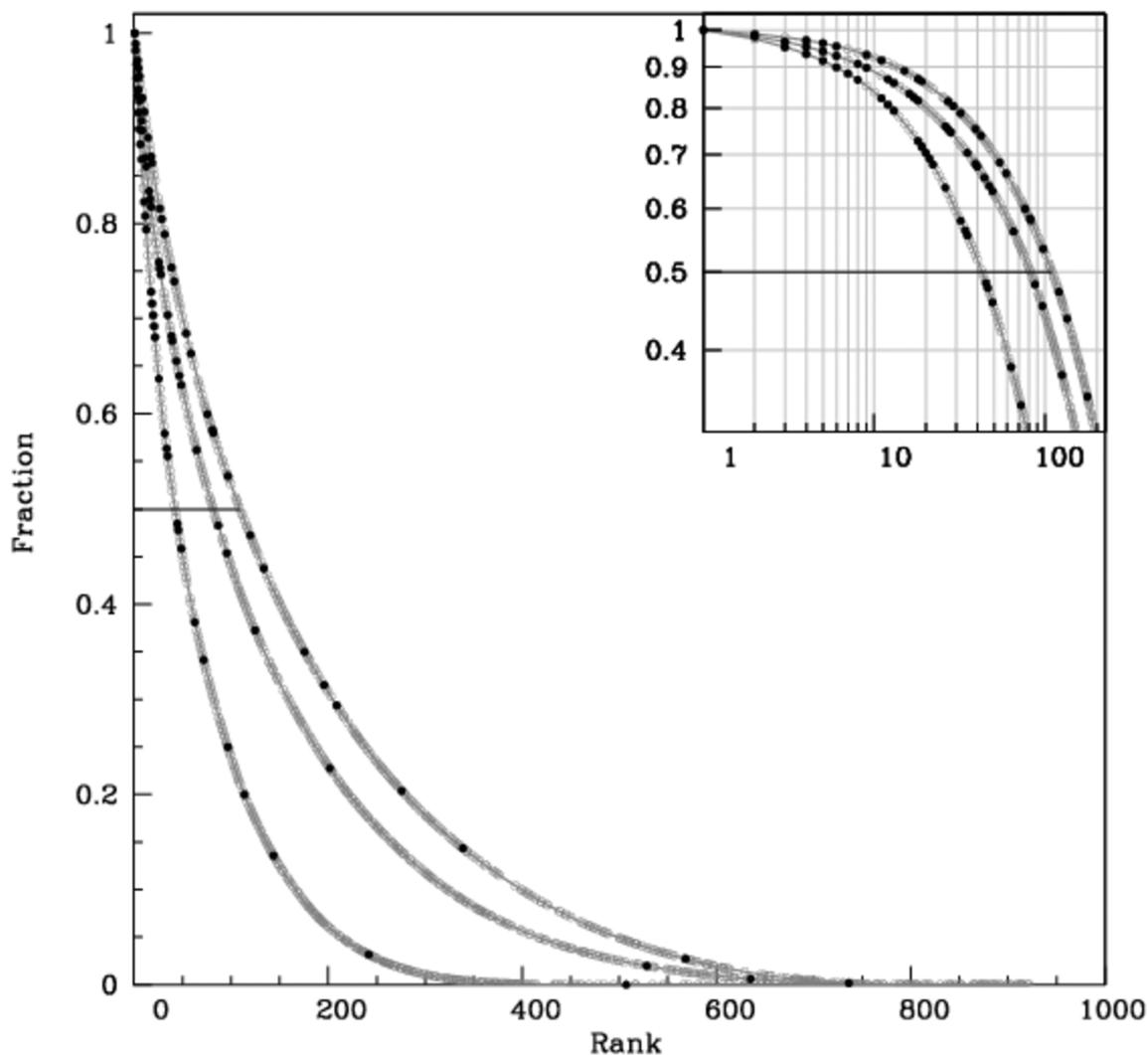

Figure 4. The cumulative fraction of Tori5 (right), Read10 (left, and Tori25 (center). Open circles are AAS members; filled circles are the honored sub-set. The horizontal bar shows half of all signal; the inset is the upper left of the plot in log units.

Figure 4 shows how productivity becomes more concentrated over time. The sum over all individuals of the three measures is normalized in each case to one, and the plots show the incremental sums for each of these measures, beginning with the lowest ranked (922) individuals score and incrementally adding the next highest ranked individual to the running sum, until, at rank 1, the entire signal is added in. The symbols represent AAS members (open circles) and honored individuals (filled circles); individuals at the same rank are different for each measure.

From this plot one can see how the scientific impact of an age-selected cohort gets in-creasingly concentrated as the cohort ages. As an example, one can read off the plot that 110 individuals (12% of 922) were responsible for 50% of the Tori5 early career measure, but it only required 43 individuals (5%) to account for 50% of the Read10 measure 27(+/- 3) years past the PhD. The Tori25 mid career sum is intermediate to the two extremes.

It is interesting to note that since 45% of the cohort had Read10 scores of zero, i.e. they are no longer authors, we find that 50% of the Read10 signal is accounted for by 10% of the authors, exactly as suggested by de Solla Price's (1963) Lottke power law argument.

**Comparison of Indicators**

All the indicators studied here can be viewed as measures of an individual's research performance, past current, and future. Were they "perfect" indicators they would correlate exactly with each other. They are not, and they do not. Using cross-comparisons we can gain insight into the strengths and weaknesses of each indicator.

*Tori25 vs. Tori5*

One of the key reasons for establishing set of bibliometric indicators is that they have predictive power; citations counts can predict prizes (Garfield and Malin 1968) and they can predict future citation counts (Hirsch 2007). By comparing Tori5 with Tori25 we can examine how well citations five years past the PhD predict citations 25 year past the PhD.

Figure 5 is complex. It shows the Tori score 25 years past the PhD minus the score 5 years past the PhD versus the score expected from the Tori5 score alone. This essentially shows the citations accrued between 5 and 25 years past the PhD compared with what would be expected from an extrapolation of the counts at 5 years past the PhD. We note explicitly that these counts include citations made between 5 and 25 years past the PhD to articles that were written before five years past the PhD.

For the expectation value of (Tori25 - Tori5) we simply multiplied Tori5 by 10.47. This was derived by using the K05b model for an individual's research performance, which assumes continuous production of equal impact articles over a lifetime, taking into account the measured obsolescence of articles with time. This is different (smaller) than would be obtained by the simple assumption (Hirsch 2005, Pepe & Kurtz 2012) that citations increase quadratically with age by about 30%.

The symbols in the diagram represent three activity levels. We chose activity based on number of publications in the five-year period 1997-2001; we make no correction for co-authorship, or type of publication (we include conference abstracts, as well as refereed full journal articles). The filled black dots represent persons who are authors of more than twenty publications in the five-year period. Open circles are authors on one to twenty publications, and the small gray dots are individuals who were not an author on any astronomy publication during the period.

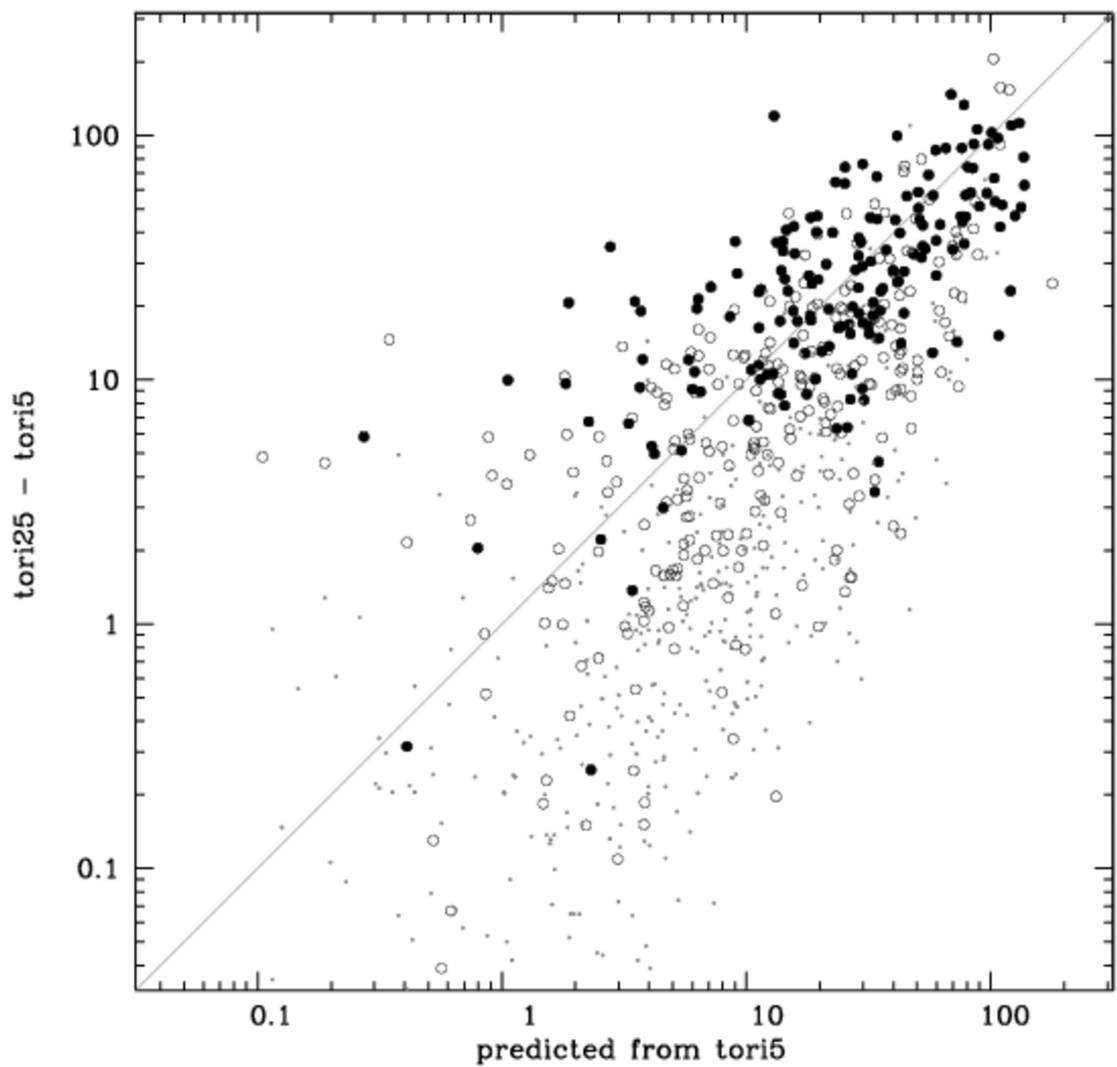

Figure 5. The additional Tori citation measure obtained in the 20 years between 5 and 25 years past the PhD vs. the expectation based on the 5 year measure. Filled circles are individuals with more than 20 publications between 1997 and 2002, open circles are individuals with 1 to 20 publications during that time, and the dots are individuals with no publications in that time.

Looking at figure 5 we see that the different activity classes show different behavior. The high activity individuals (black dots) lie toward the upper right of the diagram (the high citation end) and essentially follow the prediction, with a small offset (in the direction of being less productive than the prediction) and with a scatter of a multiplicative factor of two.

The mid-low activity individuals (open circles) have a nexus that is lower in actual Tori and with a prediction that is systematically higher compared with the actual than the higher ac-tivity individuals. The zero activity individuals (small gray dots) show a similar, but more pronounced pattern; this is the pattern of individuals who once did research in astrophysics, but have at some point stopped.

The mid-low activity individuals are likely people who have non-research responsibilities, but who are still involved in astronomical research in varying levels. Figure 6 shows the relation between activity level and the ratio of the predicted citation impact divided by the actual measured value.

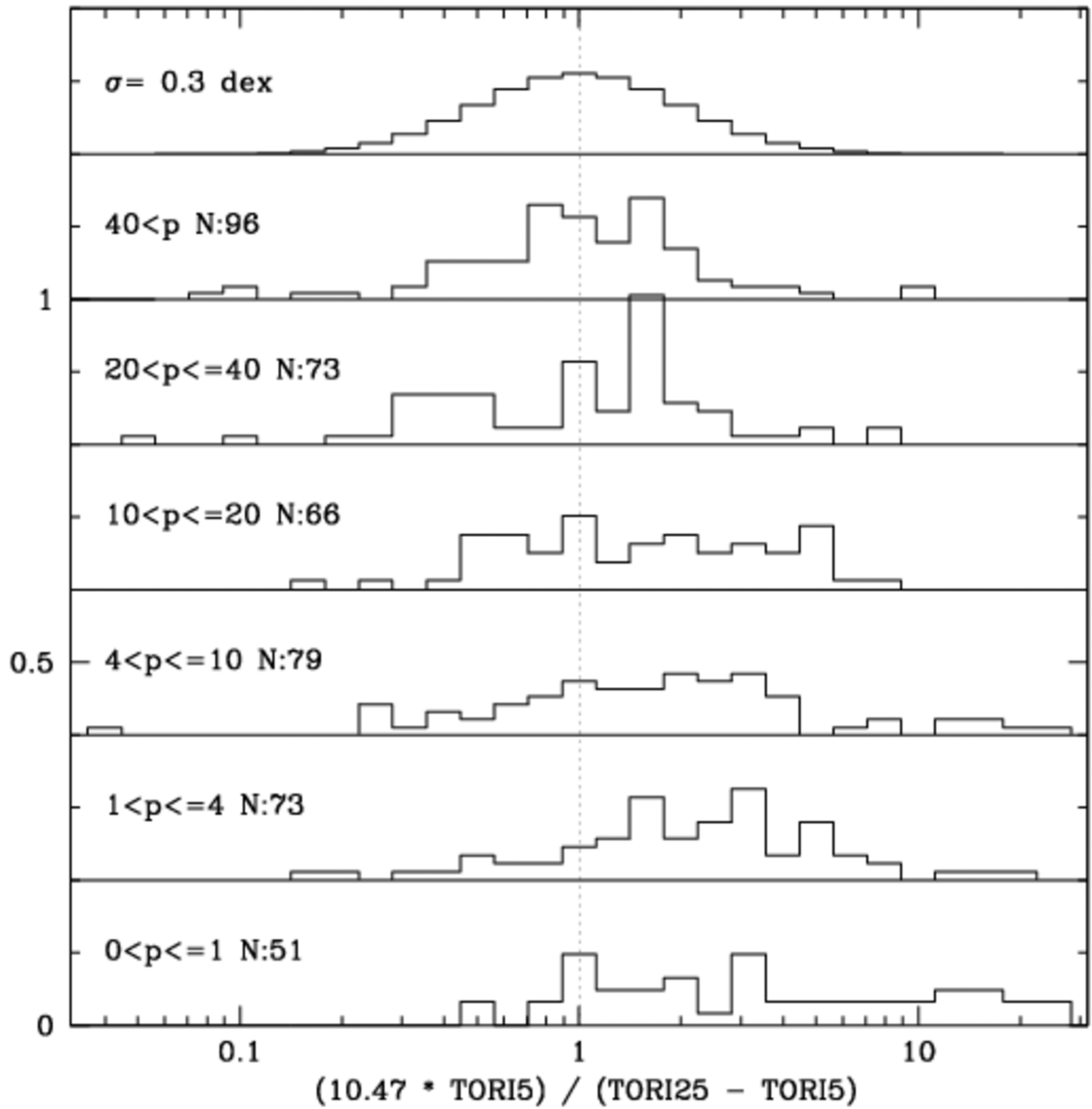

Figure 6. The distribution of the new tori measure vs the expectation based on the 5 year measure, as a function of p, the number of papers authored in 1997-2002. The top is a log-normal distribution with $\sigma = 0.3$ dex, to guide the eye.

Figure 6 shows the distribution, in equal logarithmic bins, about the predicted values of Tori25 for different activity levels (the number of papers in the 5 year period: hereafter p); a value of one means the measured value is the same as the prediction.

At the top is a normal distribution, with a standard deviation of 0.3 dex (a factor of two); this is not a fit to the data, but is meant to guide the eye. Below we plot the distribution for six different activity levels, as marked on the plot. Notice that at the highest activity level (more than 40 publications in the five year period) the distribution is symmetric about the predicted value, and is consistent with being drawn from a normal distribution (using the Kolmogorov-Smirnov test). Next notice that as the activity levels decrease the distributions flatten out and shift to the right (where individuals underperformed the prediction).

We can draw a number of conclusions from this plot. First is simply that the relevant measure is the logarithm of the citation measure, not the measure itself. That the distribution is lognormal (Shockley 1957; Mitzenmacher 2003) has important implications for any practical use of these measures.

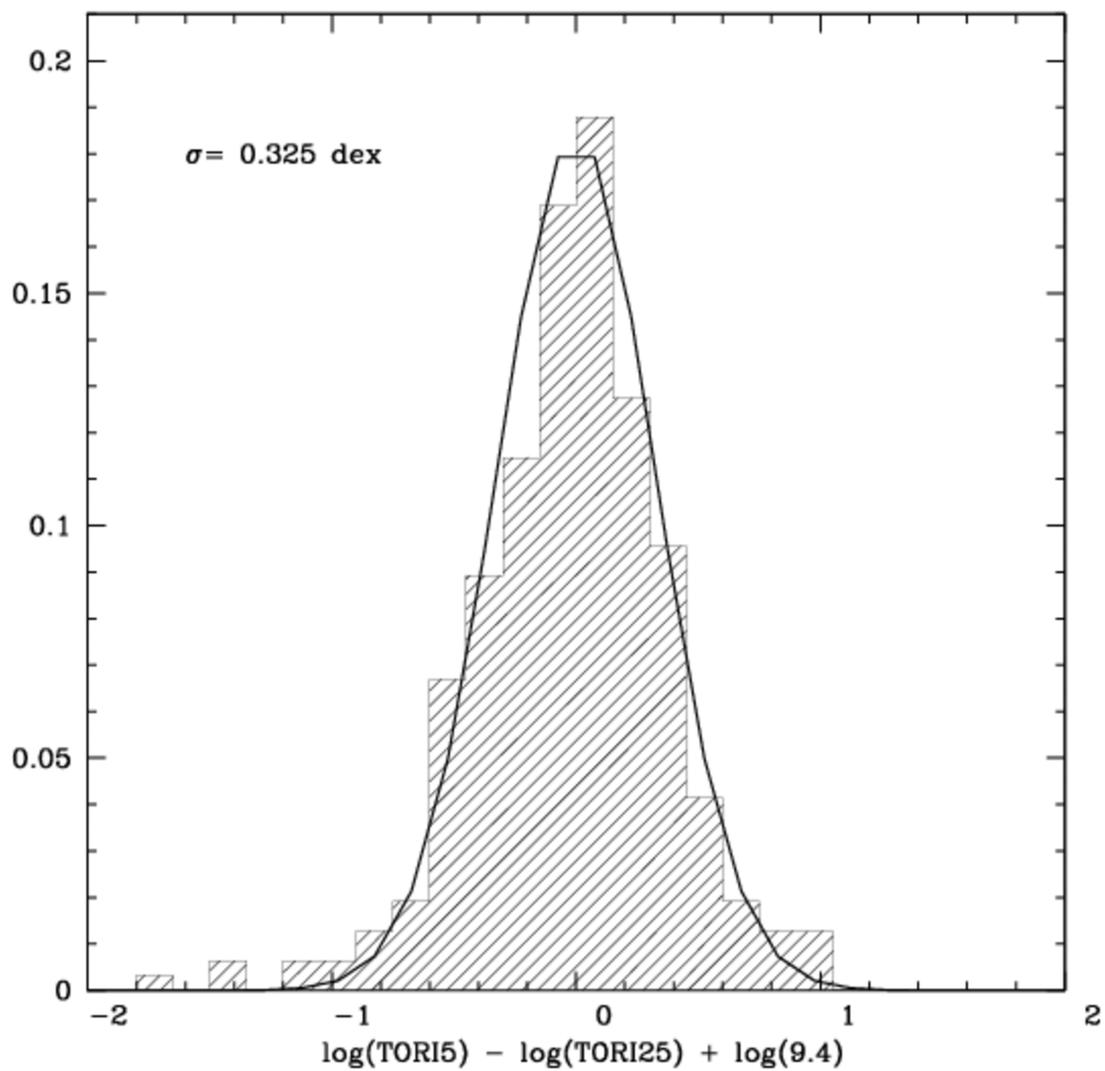

Figure 7. The distribution of the difference between tori5 and tori25 for individuals with 5 or more publications in the 1997-2002 period. The log-normal distribution with σ = 0.325 is to guide the eye.

| **The active astronomer (AA) selection** |
|---|
| |
| The active astronomer (AA) sample is chosen to be persons whose work has had a measurable impact over the entire 40 years of the study. Also they are chosen to be pure astronomers (not planetary scientists) and we have further removed eleven individuals who have well cited, but not downloadable works (reference books, software, and scientific instruments). The sample contains 184 active astronomers, all, or almost all, of whom have passed the peer review necessary to have been granted tenure, or its functional equivalent. |
| The exact criteria for inclusion in the AA sample is:<br>    1. At least 1 citation 5 years past PhD<br>    2. Author on at least five astronomy research publications in 1997-2001<br>    3. Papers written between 1991-2001 downloaded at least once during 2000-2001<br>    4. Papers written between 1991-2001 downloaded at least once during 2012-2013<br>    5. Papers written between 1991-2001 have in 2014 a combined Tori index greater than that of the median single ten-year-old paper (0.16).<br>    6. Fewer than one in twenty papers appears in a planetary science journal<br>    7. Is not one of the eleven individuals with well cited non-downloadable works |
| This is obviously a very strong *a posteriori* selection; it is intended to find the minimum variation in the measured quantities, after removing all known systematics with 20/20 hindsight. Relaxing any of the restrictions results in increased variance for all measures; tightening these restrictions further has little effect. |

Table 1. Definition of the AA sample.

      Second, the spread of the distributions is large. Figure 7 shows the sum of the top four histograms in figure 6 (with p > 4; 314 individuals) against a Gaussian distribution with standard deviation of 0.325 dex. This is a strongly biased, *a posteriori* selection, only choosing individuals who, between 20 and 25 years past the PhD, were authors of at least one paper per year (about a third of the original sample). Even with this sample the ability, at 1-$\sigma$ to predict a person's future citations based on their citations at about the time of being hired as assistant professor, is no better than a factor of two. This also has important implications for any practical use of these measures.

      It is also possible to measure the predictions between different times. We here add Tori12, the citation measure 12 years past the PhD, approximately when tenure decisions are made. The results, here using the Active Astronomers (AA) sample (see box 1), and computing the sample mean and standard deviations:

$\log(\text{Tori25}) = \log(\text{Tori5}) + 0.961 +/- 0.332$ (1-$\sigma$)

$\log(\text{Tori12}) = \log(\text{Tori5}) + 0.539 +/- 0.243$ (1-$\sigma$)

$\log(\text{Tori25}) = \log(\text{Tori12}) + 0.423 +/- 0.162$ (1-$\sigma$)

      By choosing a sample of individuals with a career in research, and by using a cumulative

citation measure, where some of the signal at the latter measurement point is due to citations to papers that had already been published at the start point, we measure a best-case scenario for the prediction of an individual's achievement.

We can also look at the number of additional citations a person receives, for example how well can we predict the number of new citations (including to already published articles) one will receive between 12 and 25 years past the PhD, given the citation count at 12 years past?

$$\log(Tori25-Tori12) = \log(Tori12) + 0.185 +/- 0.272 \ (1\text{-}\sigma)$$

Notice that the scatter is much larger, 0.272 dex (1-$\sigma$), a multiplicative factor of 1.87.

*Downloads*

Reads (downloads) provide an alternative method for measuring impact (Kurtz & Bollen 2010). Reads and cites have very different statistical and systematic properties. While all citations are created by authors, not all paper downloads are caused by authors; indeed there are several classes of readers and they read different things (Kurtz & Bollen 2010). This makes downloads a intrinsically more noisy signal than citations, which are all intentional. In the present study the primary downloads data were taken in 2000-2001, before the ADS became popular outside of professional astronomers; and astronomy is a field that has, essentially, no practitioners, all astronomers are researcher/authors.

We also use downloads data from 2012 and 2013; We identify user/reader types by usage pattern and filter the logs accordingly. This way we select only those individual readers who have usage patterns consistent with being professional researchers.

We do have the problem of differential usage leak (Moed and Halaki 2015). While in 2001 astronomers used the ADS nearly universally, this was not true for planetary scientists. Figure 8 shows the relation between the Read10 statistic and the Tori citation measure for exactly the same 10 year span of papers (Tori10/14), but measured in 2014. The open circles represent individuals for whom 5% of more of their papers in the 10 year period appeared in pure planetary science venues. The relation of reads to cites for these individuals is obviously quite different from the main group, in that their papers received substantially fewer downloads via ADS than similarly cited pure astronomy papers. The AA sample (box 1) excludes them.

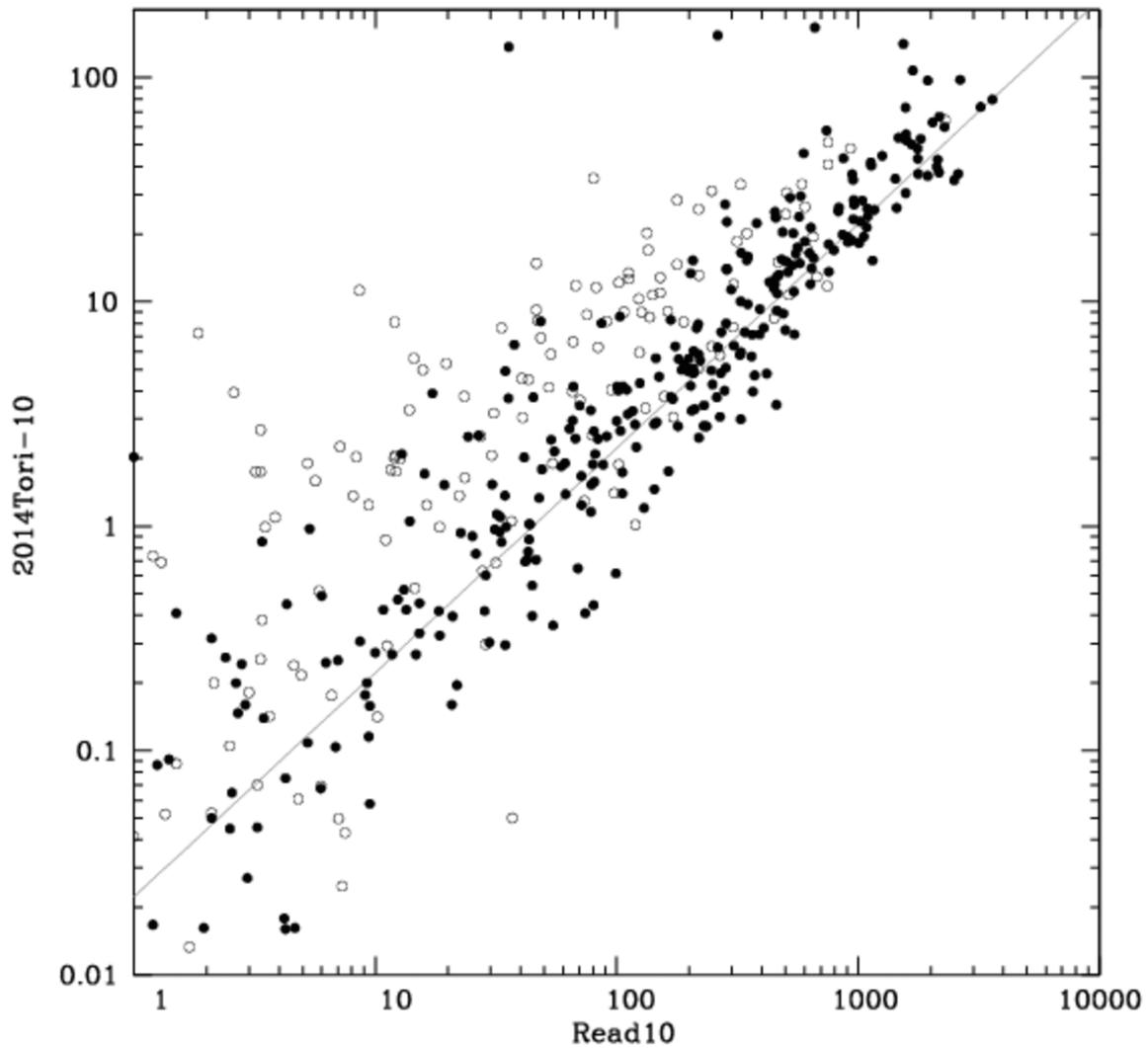

Figure 8. The distribution of the Read10 download statistic, measured in 2002 for papers written after 1991 vs tori10/14 the tori citation statistic for the exact same sets of papers, measured 12 years later, in 2014. The open circles represent individuals who published at least one paper in the planetary science literature during that time; the filled circles did not. Note the clear separation between planetary scientists and other astronomers.

Again fitting a normal distribution to the log of the ratio of Tori10/14 with Read10 for the AA sample we find:

log(Tori10/14) = log(Read10) + 1.63 +/- 0.199 (1-σ)

The scatter in this relation, 0.199 dex, is a multiplicative factor of 1.58. This suggests that evaluations of individuals based on download statistics will have different results than evaluations based on citation statistics, even when over exactly the same set of papers with nearly all readers being also authors.

In what follows we continue to use the AA sample. We only investigate the relations between different citation (Tori) measures and download (Read10, etc) measures for career research astronomers, removing all known systematics.

When comparing the productivity of individuals, using different indicators computed at different times, it is necessary to be able to differentiate between effects caused by individuals changing their productivities, and changes solely due to measurement systematics. We examine this issue by looking at several pairs of similar indicators.

Figure 8 shows two different indicators (Tori10/14 and Read10), taken at different times over the exact same documents. The scatter (for the AA sample) about the linear relation, 0.199 dex, is a direct measure of the systematic and statistical variation of these indicators.

Figure 9 shows the same indicator, also taken at different times. Read10/14 measures the downloads by scientists during the years 2012 and 2013 of exactly the same sets of papers used in the Read10 measure, which covered downloads by scientists in 2000 and 2001. For the AA sample the relation is:

log(Read10) = log(Read10/14) + 0.179 +/- 0.167 (1-σ)

Notice that the scatter in this relation 0.167 dex is smaller/similar to the comparison of Read10 with the current citation measure for the Read10 documents (Tori10/14).

Read10 is a measure of current activity, and includes some documents that were published during the measurement period. The quickly decaying current awareness mode (K05b, Kurtz & Bollen 2010) does not have the same relation with citations, and likely also downloads, as later occurring readership modes. Also Read10 counts downloads of grey literature documents, which also have different relations with later citations and/or downloads than standard journal articles. These factors combine to add additional systematic scatter into any relation using Read10.

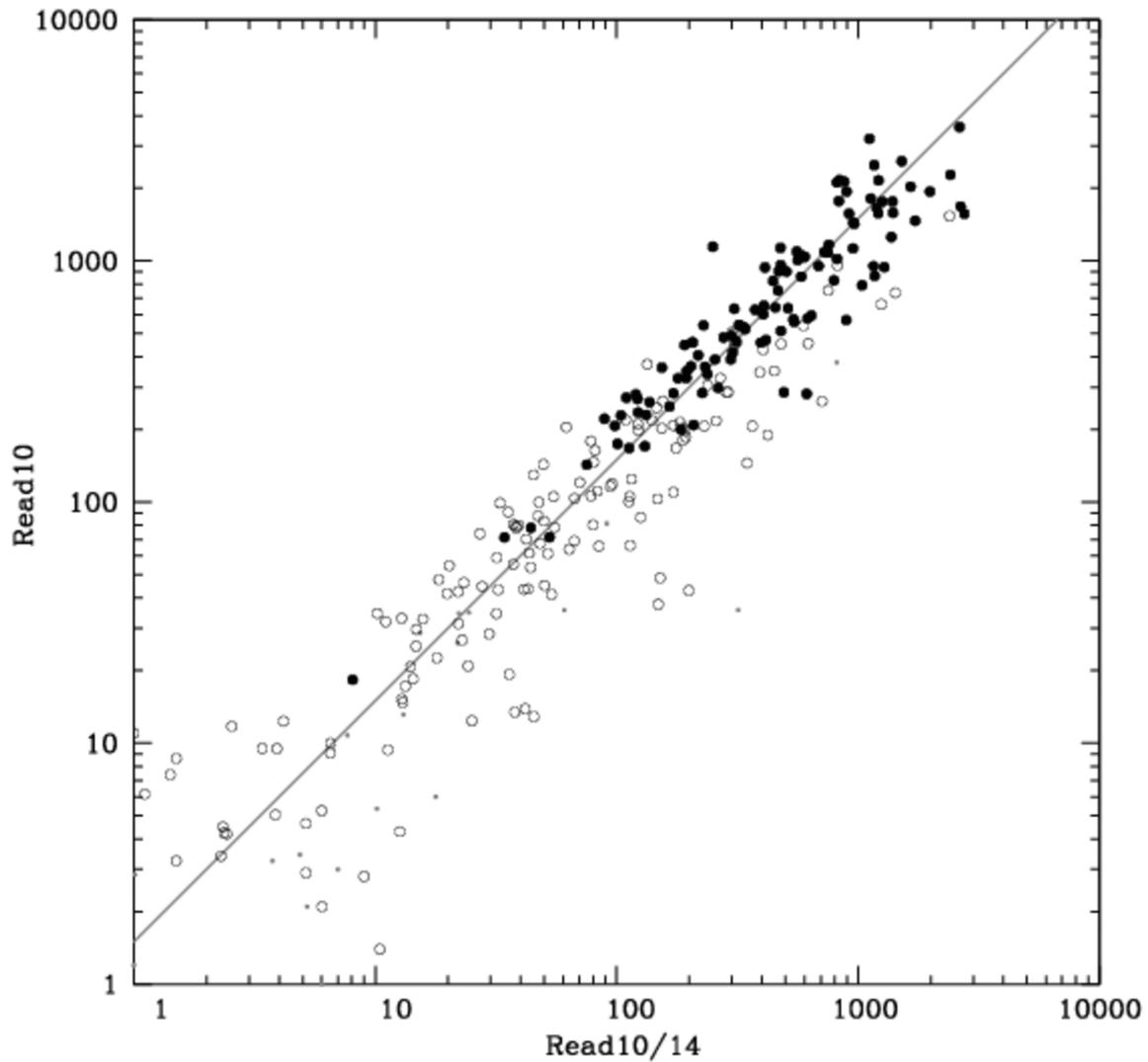

Figure 9. Read 10 vs Read10/14 an identical measure over the exact same sets of papers, but measured using the 2012-2013 reads. Symbols are the same as in figure 5.

How large is this systematic? We first develop two new measures which are hardly affected by systematics in Read10: RQuot and RQuot14. RQuot is the number of normalized reads during the years 2000-2001 to an individual's papers written before 1991. RQuot14 is the number of normalized reads to these same papers during the years 2012-2013.

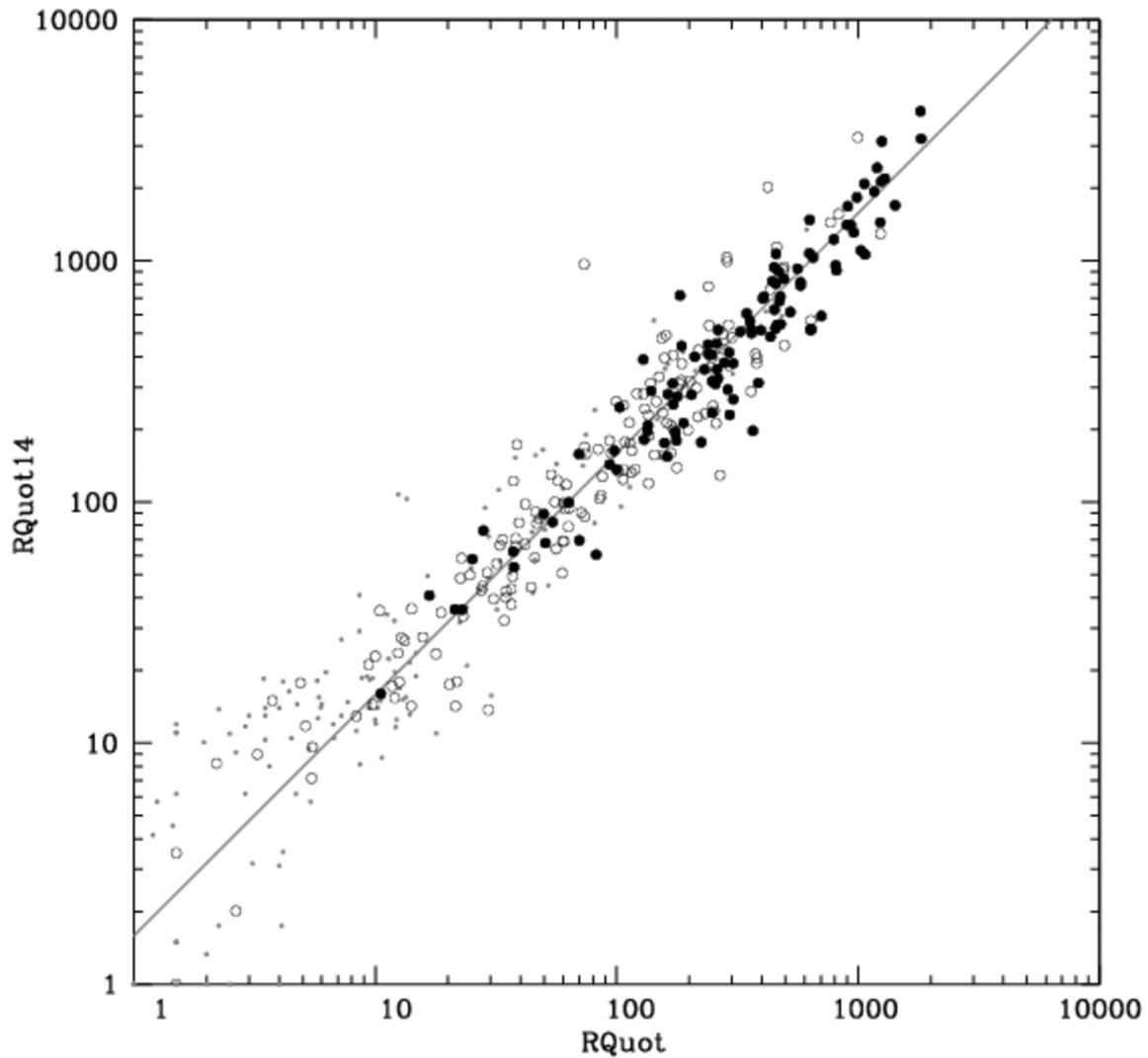

Figure 10. RQuot, a normalized download statistic aggregating by astronomer papers written before 1991, and measured using the 2000-2001 reads vs RQuot14, a similar statistic, using exactly the same sets of papers, measured using the 2012-2013 reads. Symbols have the same meaning as in figure 5.

Figure 10 shows the relation between RQuot and RQuot14, similar indicators over identical sets of papers measured at two different times. For the AA sample the relation is:

log(RQuot) = log(RQuot14) - 0.179 +/- 0.150 (1-$\sigma$)

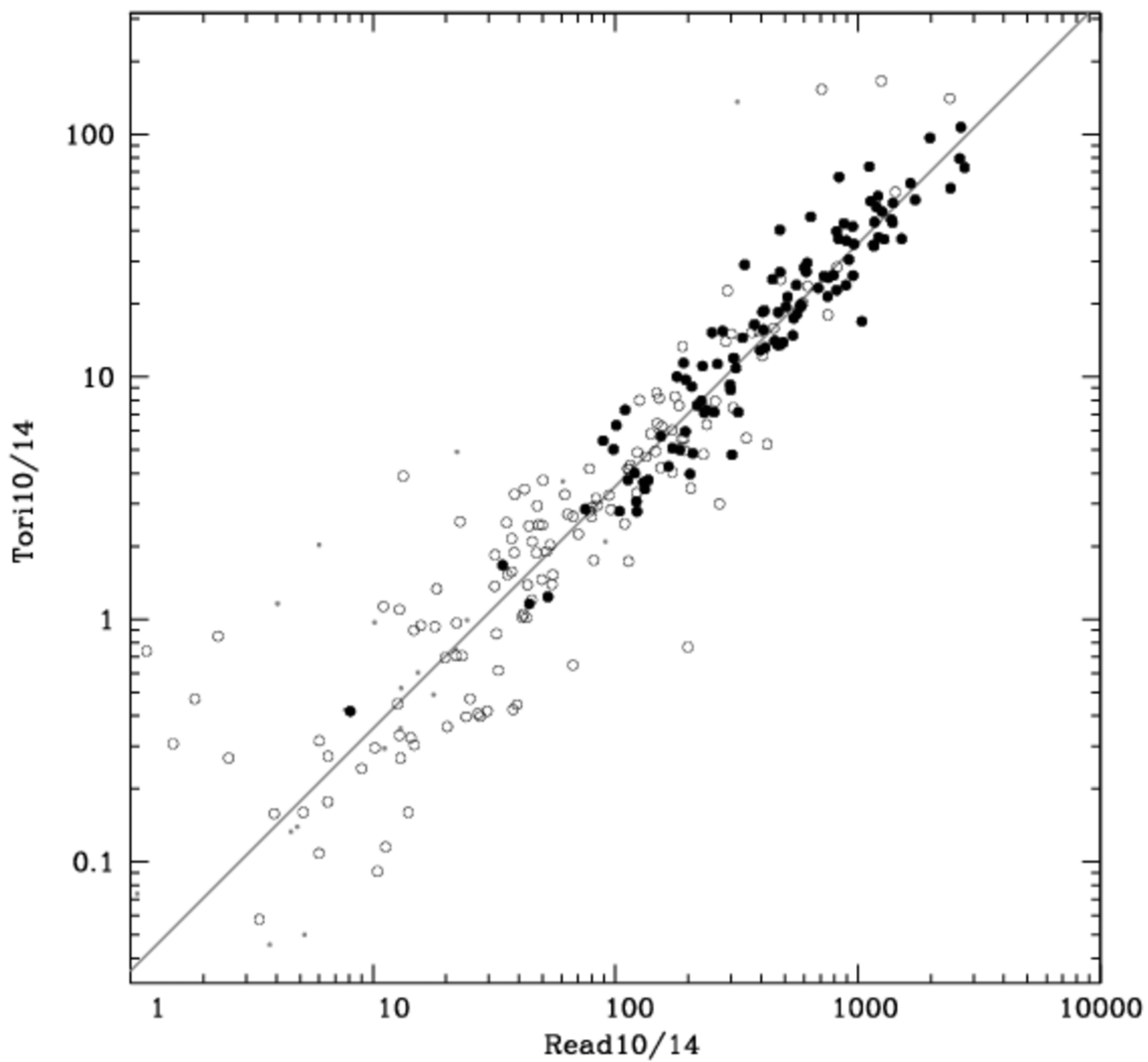

Figure 11. Read10/14 vs Tori10/14. Symbols have the same meaning as in figure 5.

Figure 11 shows the relation between Tori10/14 and Read10/14, different indicators over identical sets of papers measured at (about) the same time. For the AA sample the relation is:

log(Tori10/14) = log(Read10/14) + 1.45 +/- 0.165 (1-σ)

Again notice that the scatter in these two comparisons is similar: 0.150/0.165 dex (1-σ); this is smaller than for the comparisons involving the log of Read10. From the four comparisons we can conclude that the (1-σ) scatter in each individual measurement of an individual author is 0.11-0.14 dex, with Read10 and Tori10/14 having similar but somewhat larger scatter.

We can see from the previous four plots and equations that both the citation statistics (Tori) and the download statistics (Read10, etc) are stable over time, and that the read statistics and the citation statistics are, within the scatter, measuring the same thing.  This puts very strong constraints on any theory of citation (Nicolaisen 2007; Bornmann and Daniel 2008).  The linear relation between Read10 and Tori10/14 shows that downloads do, indeed, predict citations, which agrees with the Moed and Halawi(2015) critique of K05b.  The per article effects which caused K05b to suggest otherwise produced the increased scatter in Read10, when averaged over individuals.

*Comparing Astronomers Early and Mid-Careers*

Predicting future performance from contemporary measures is one of the principal goals of individual evaluations of individuals, whether bibliometric or not.  Figure 5 showed the relation between Tori5 and Tori25, showing how well an early performance measure can predict a later one.

If everyone always performed at the same rate early measures could perfectly predict later ones (ignoring measurement scatter).  They do not; people change over time.  We here characterize and describe the change, as seen by comparing three bibliometric measures, Tori5, Tori25, and Read10.  Each of these measures represents a different time in an individual's career.

Citation measures are cumulative. For an individual who has constant production of equally well cited papers, more than half the signal of any citation measure is always due to papers published before the midpoint in time of the time span under consideration. The exact point is a function of the diachronous obsolescence relation.  Using the K05b obsolescence values the time when the papers representing half of the Tori25 signal had already been published is about nine years past the PhD, for Tori5 is about three years (we assume that individuals began publishing two years before the PhD).

Download measures, such as Read10, are not cumulative they are near instantaneous rates.  The most recent papers are read more than older papers, using the K05a reads obsolescence function we find that half the signal in Read10 comes from papers that are three years old or younger.

Figure 5, Tori25 vs. Tori5, effectively shows estimates of an individual's relative performance at three and nine years past the PhD, a six-year difference.  This points out one of the principal difficulties in measuring individuals with citation measures:  the measures are always weighted towards the first half of the individual's career.

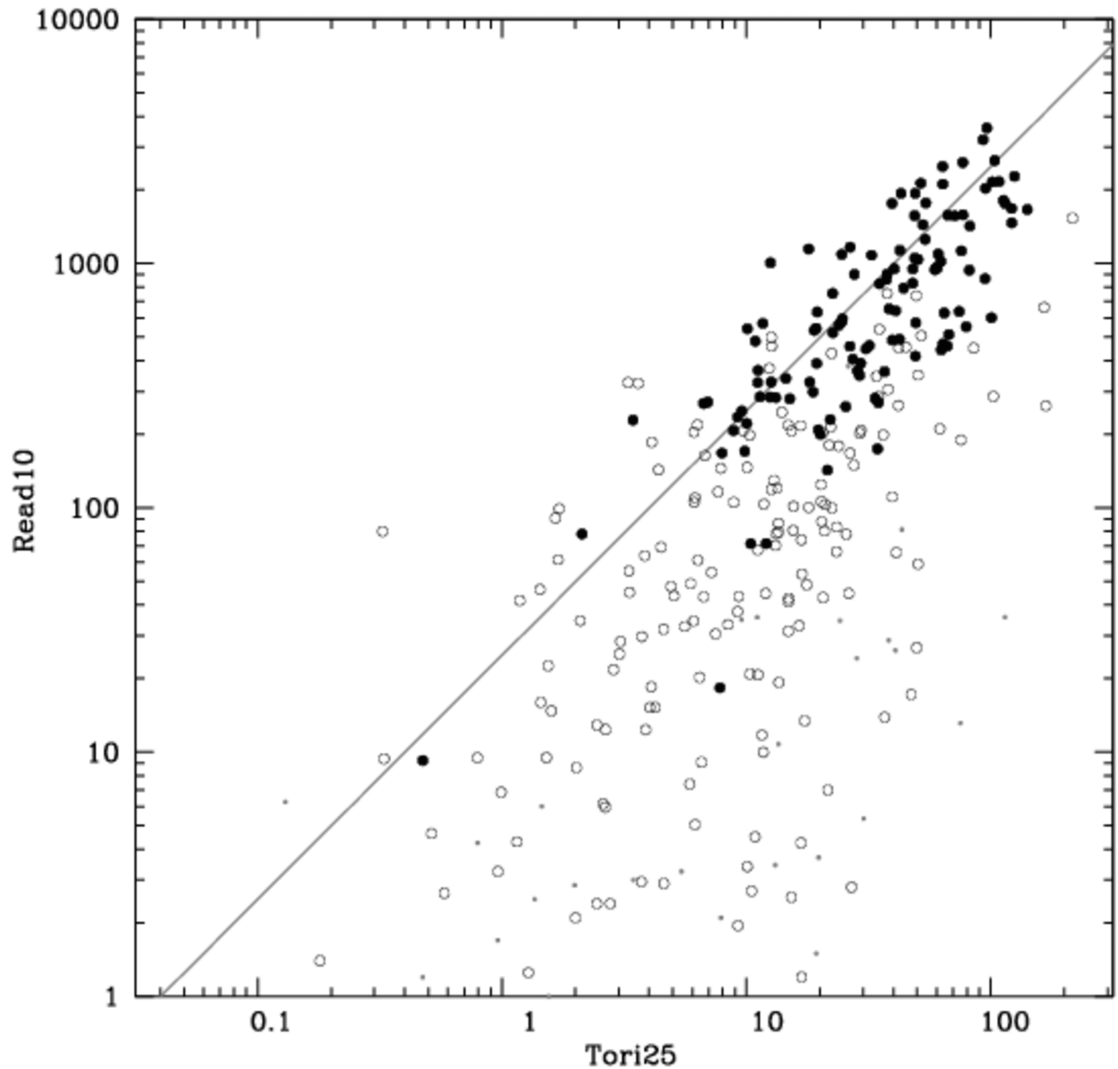

Figure 12. The relation between two contemporaneous measures of individual performance, Tori25 and Read10. Symbols have the same meaning as in figure 5.

Downloads have the opposite problem, they principally measure recent performance; Read10 is especially designed to have this property. Figure 12 shows Tori25 plotted against Read10, showing differences in productivity measured at effective date 9 years past the PhD and 24 +/- 2 years past the PhD, an effective time span of 15 years.

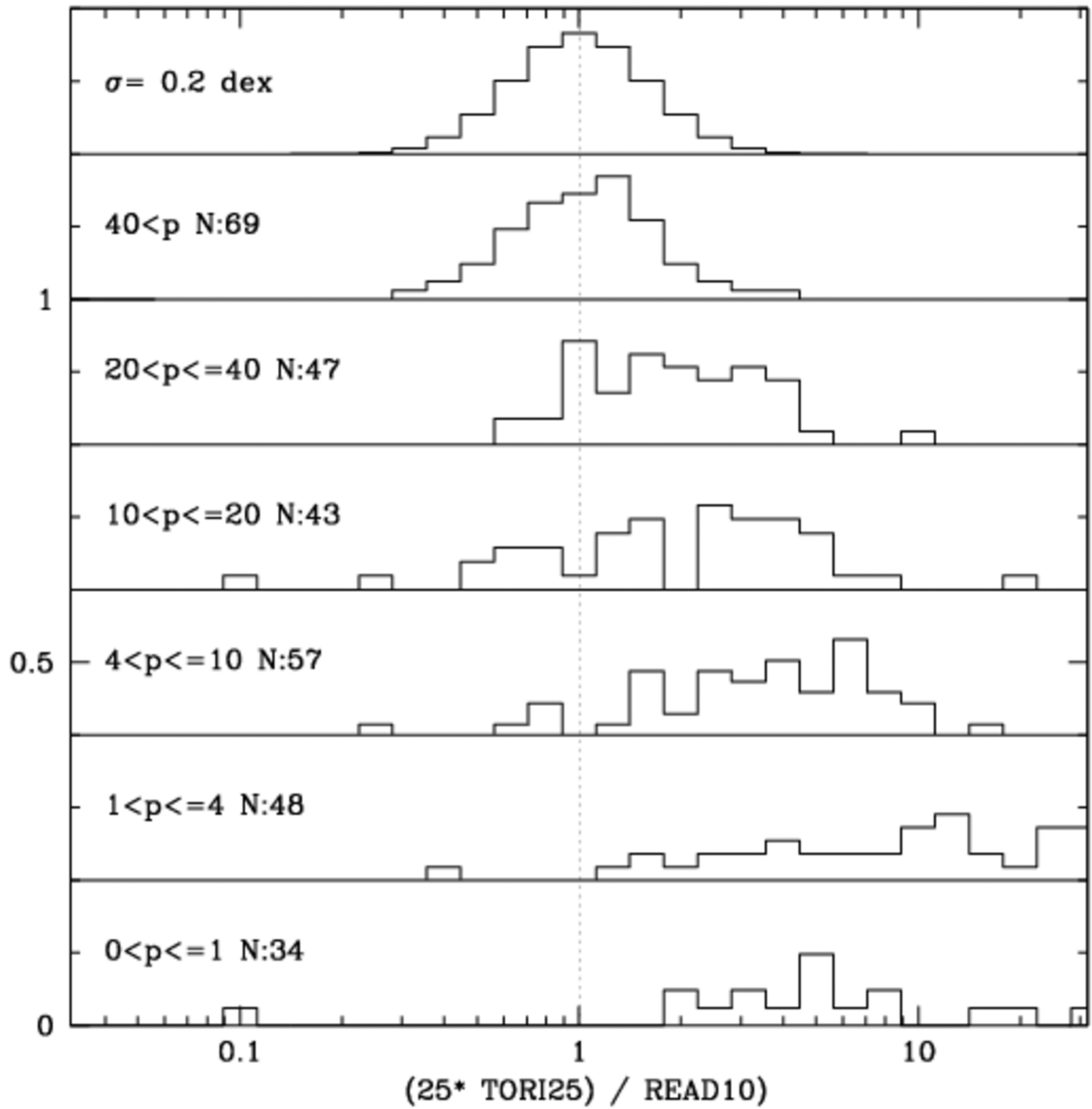

Figure 13. Another view of the relation of Tori25 with Read10. The top histogram shows a log-normal distribution with a $\sigma = 0.20$ dex, to guide the eye. Note that only the most prolific individuals can be approximated by that distribution. This figure has the same format as figure 6.

Even for the AA sample this distribution cannot be described by a normal distribution; a K-S test rejects that hypothesis. We report a value for the sample standard deviation of the comparison. The extra uncertainty is indicated by the ~ symbol in the equation, replacing the equal sign. Figure 13, which has the same format as figure 6, shows what is happening. As the current activity measure (p, the number of publications authored in the 5 years 1997-2001) decreases the current productivity measure, Read10, decreases compared to what would be expected from the earlier productivity measure, Tori25. As noted above, figure 5 also shows this, but to a lesser degree.

log(Tori25) ~ log(Read10) - 1.18 +/- 0.347 (1-σ)

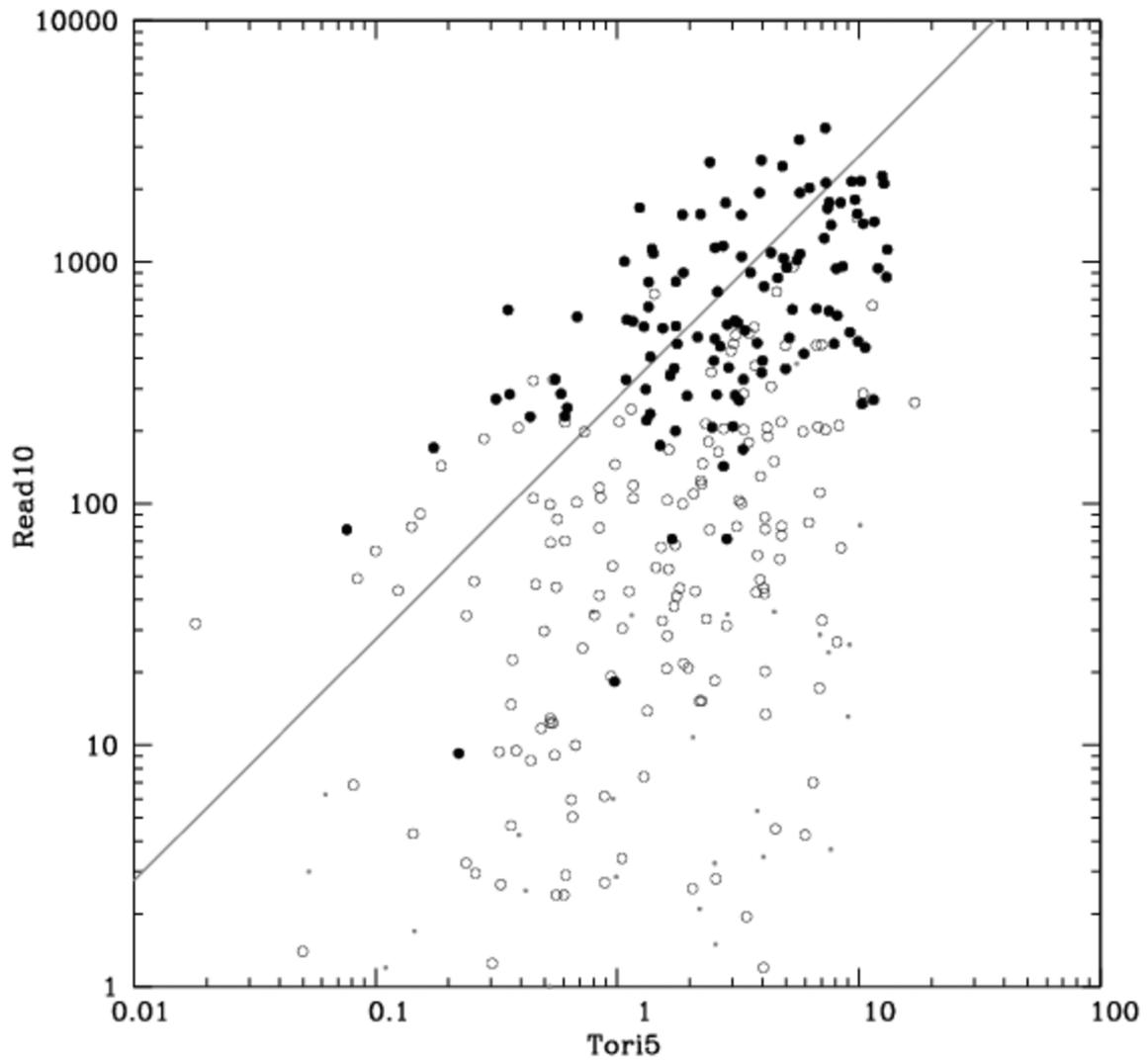

Figure 14. How well early citations match mid-career performance. Tori5 vs Read10. Symbols have the same meaning as in figure 5.

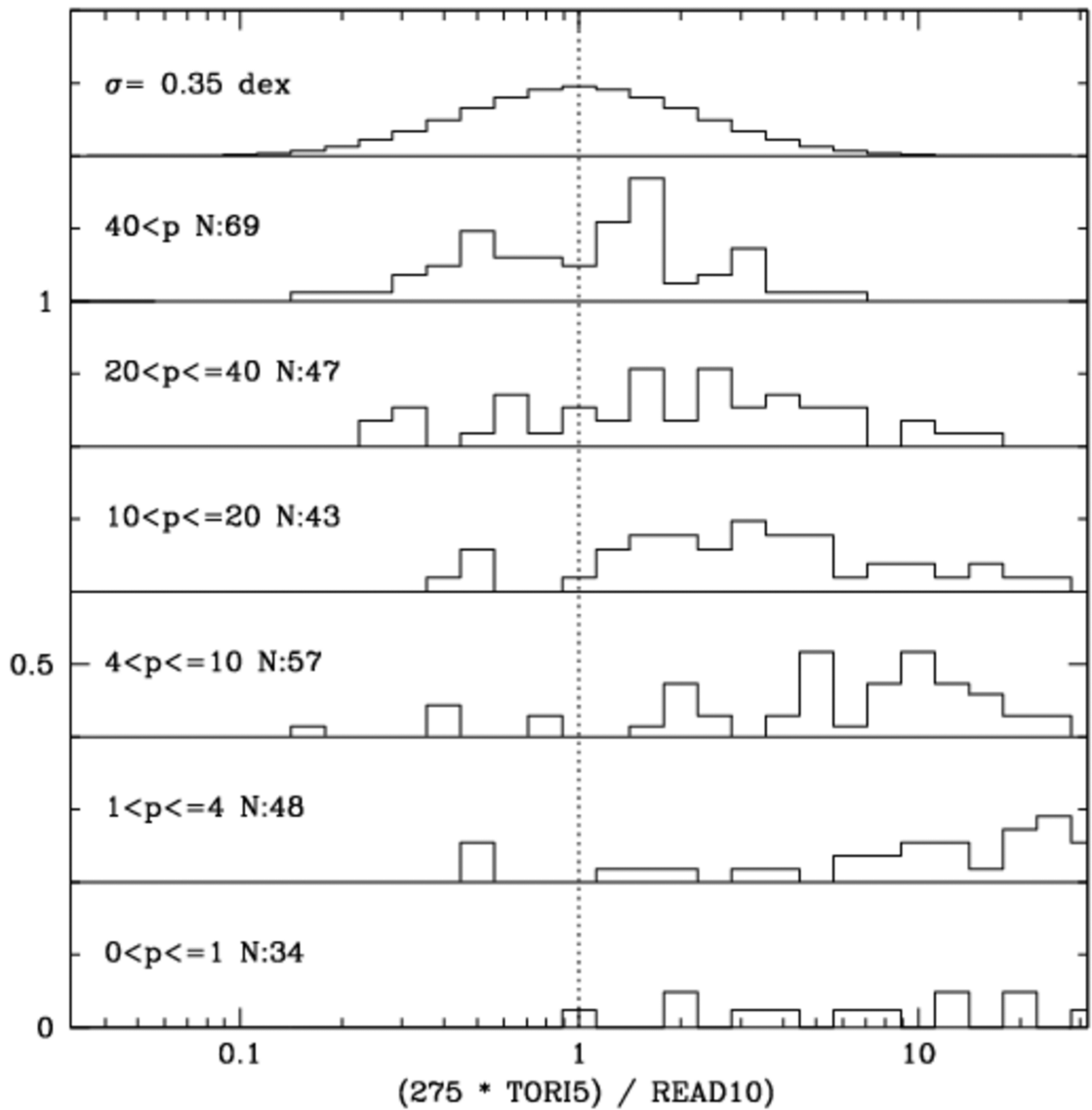

Figure 15. Another view of Tori5 vs Read10. The histogram on top is a log-normal distribution with a σ of 0.35 dex. The format of the figure is the same as figure 6.

Figures 14 and 15, Tori5 vs Read10, show this effect even more strongly. In approximate terms figure 15 shows the difference between one's measured productivity at age 50 and what would be predicted based on one's productivity at age 30; separated into groupings by age 50 activity. Again, even for the AA sample, this is not a normal distribution, and we report the sample mean and standard deviation using the ~.

log(Tori5) ~ log(Read10) - 2.14 +/- 0.485 (1-σ)

Aside from the fact that some individuals become less productive over time, and that this effect is correlated with their activity, figure 14 also shows an additional difficulty with using early cita-tion measures to characterize individuals. In figure 14 there are individuals who form a locus above and to the left of the arbitrary linear relation drawn on the plot; they have substantially higher Read10 scores than Tori5 would predict.

While there are "late bloomers" who increase their productivity as they age, most of the effect seen is likely due to differences in doctoral programs. It is quite common for PhDs to be granted after four years of graduate school, and after eight years is also not uncommon. Some individuals write many papers as grad students, and some write their first after they receive their degrees. As citations increase quadratically with time, standardizing age to time since the PhD can easily produce systematic differences of 0.3 dex (a factor of two).

**Tori5 as a predictor of long-term productivity**

As can be seen in figure 14 Tori5, the citation measure five years past the PhD has substantial scatter and systematics. Figure 11 suggests that contemporaneous download data on the same individuals would suffer similar difficulties (download data for the cohort studied here does not exist for the relevant time period). Despite these problems, how well does Tori5 (and by extension early download measures) actually predict future productivity of individuals and how does it compare with human judgments?

Figure 16(a-d) shows the same data as figure 14, Tori5 vs. Read10. Here we bin the data to show how well a knowable measure of performance as a student and post-doc (Tori5) can predict mid-career performance (Read10). The four sub-figures show the comparisons with Read10 for four Tori5 bins; (a) is for the top 10% ($90^{th}$-$100^{th}$ percentile). (b) is for the $2^{nd}$ 10% ($80^{th}$-$90^{th}$ percentile), (c) is for the next 30% ($50^{th}$-$80^{th}$ percentile) and (d) is for the bottom half ($0^{th}$-$50^{th}$ percentile).

Each sub-figure shows, for the relevant Tori selected members of the cohort, the relative population, compared with a random selection, of binned Read10 levels. The Read10 bins are 100-$90^{th}$ percentile, 90-$80^{th}$ percentile, 80-$70^{th}$ percentile, 70-$60^{th}$ percentile, 60-$46^{th}$ percentile, and 46-$0^{th}$ percentile. All individuals in the last bin (0-$46^{th}$ percentile) have Read10 scores identically zero; we expanded the $2^{nd}$ to last bin to include all remaining non-zero scores. In terms of the fraction of reads in each bin, the 90-$100^{th}$ %-tile bin has 74% of all reads, 80-$90^{th}$ 18.5%, 70-$80^{th}$ 5.8%, 60-$70^{th}$ 1.5% and 46-$60^{th}$ 0.2%.

Scores of 1.0 correspond to what would be expected from a random selection from all 922 individuals. Figure 16(a) shows that an individual in the highest Tori5 decile is 4.3 times more likely to be in the top decile of Read10 than a random selection would be, and that that individual was substantially less likely than random to be a poor performer or to have stopped doing research.

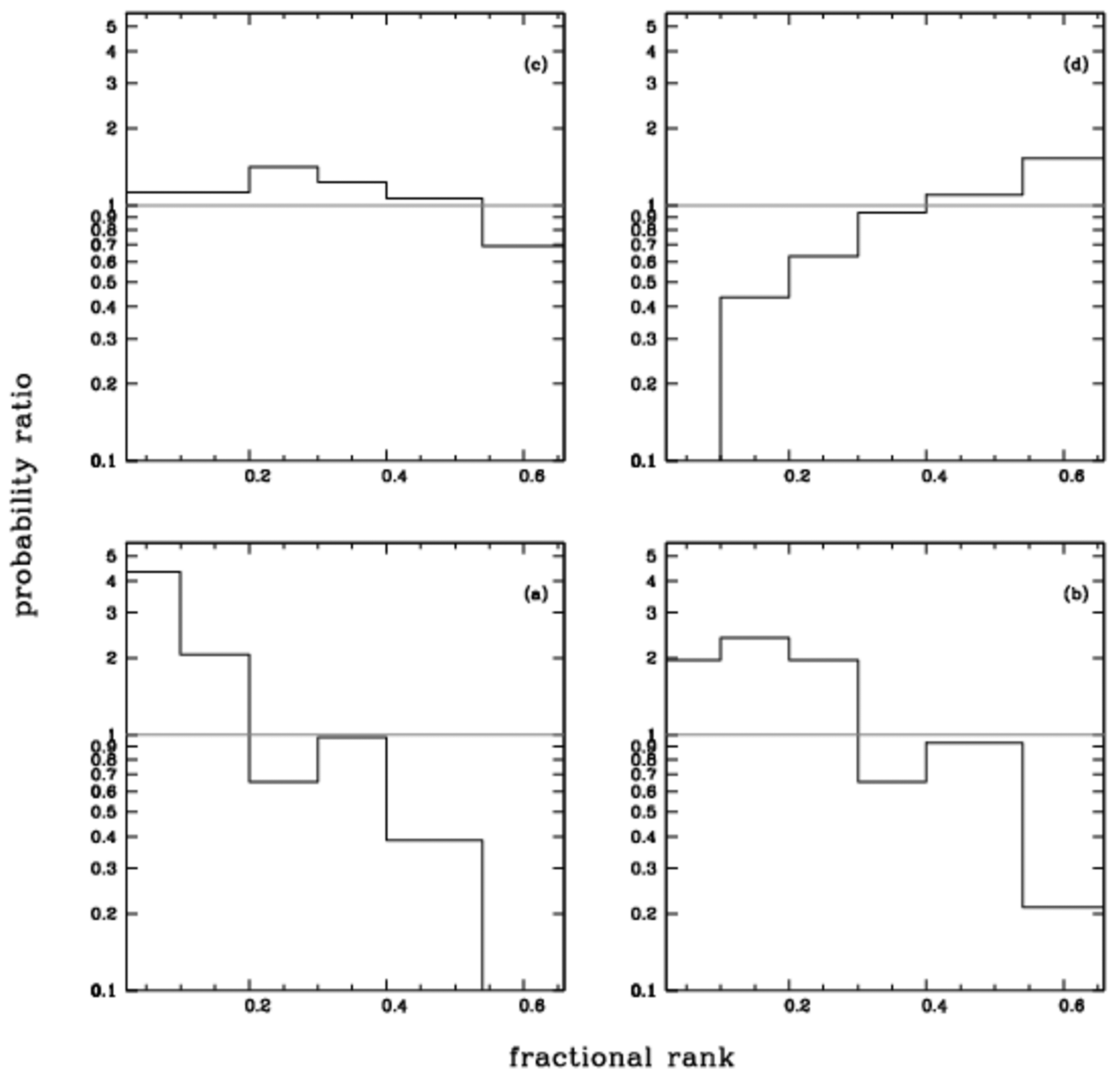

Figure 16. Yet another view of how well Tori5 can predict Read10. The four subfigures show the distribution of Read10 rank for different selections in Tori5. (a) shows the relative probability of being in different bins in Read10 (compared to one for a random selection) for individuals in the top 10% of the Tori5 distribution (the 90-100th percentiles), the first 4 Read10 bins are each 10 percentile (90-100; 80-90, 70-80 and 60-70) the fifth bin is 46-60 percentile and the final bin is 0-46, and is all individuals with a Read10 score of zero. (b) contains the 80-90th percentile in Tori5, (c) the 50-80th percentile in Tori5, and (d) the bottom half of the Tori5 distribution.

Figure 16(b) shows that the 2nd decile in Tori5 is not as steep as the 1st decile, but still gives a reasonable discrimination of high versus low producers. Figure 16(c), the 50-80th percentile group, shows very little difference compared with random, save that these individuals are somewhat less likely to leave astronomy research. Figure 16(d) shows that individuals in the lower half of theTori5 distribution are much less likely to be high producers at mid-career, and substantially more likely to have left the field entirely.

While figure 16, especially 16(a) suggests using Tori5 as an input in employment decisions at about the assistant professor level would be beneficial, it also shows that Tori5 in itself is not sufficient. Figure 16(c) shows that even people in this middle range of Tori5 become top producers. Three current members of the NAS are in this group; each is currently a very well known scientist. Each of these three individuals had a tenure track position at a major research facility five years past the PhD; clearly the expert judgment of the hiring committees saw something in them that was not yet apparent in their early citation measures.

Of course expert judgments are also imperfect (Cole, Cole and Simon 1981); to compare Tori5 with expert peer review we look at the membership of the NAS of individuals who received one of the three AAS early career (defined as five years past the PhD) prizes (Warner, Pierce and Cannon) between 1971 and 1990 (49 astronomers) and compare it with the number of NAS members in the top 49 measured with Tori5. During this time period citation counts were not as commonly used as today, so the committee decisions were likely not very much affected by them. There are 8 NAS members in the Tori5 group and 9 in the AAS prize group. This somewhat underestimates the Tori5 group, for two reasons. The AAS prizes determined the cutoff, and, while all the AAS prizewinners have had careers in astronomy research, several of those in the top Tori5 have had very successful careers in other domains, including in science administration, industry, and a well-known novelist. We therefore conclude that Tori5 performs equivalently to (but differently from) expert peer judgment.

Comparison of figure 16 with the results of Li & Agha (2015) on the relation of peer review of funding proposals with future citation outcomes suggests that both peer review and bibliometric methods are similar, in that in the mean higher scores predict mean higher scores in the future. Comparison of the large scatter seen in figures 14 and 15 with the results of Cole, Cole & Simon (1981) on peer review of funding proposals and of Olivier (2015) on peer review of scholarly articles also suggests that peer review and bibliometric methods are similar, in that both methods show large statistical scatter in their ability to predict future performance of individual researchers, research proposals, or research articles.

**Discovering Systematics**

Besides statistical errors all measures of individual performance have systematic biases. Much of the above discussion has been focused on finding and removing systematics in order to quantify the idealized core capabilities of citation and download data.

Human judgment also has systematic biases, both known and unknown. For example one hundred twenty-one men have won the Nobel Prize in physics since the last woman received it; a highly improbable occurrence absent bias. Here we use a combination of measures to discover a likely bias in the election of astronomers to the U. S. National Academy of Science.

We take the two mid-career measures Tori25 and Read10, and select the top 50 persons in each measure; 31 individuals are in both groups. We merge this list of 31 with a list of persons who have won a mid or late career prize from the AAS, or who have been elected president or vice-president of the society. Sixteen from the 31 are in the AAS list, and 15 are not.
	Now we look at NAS members.  One from the set of 15 non-AAS prizewinners is in the NAS, this person is a university professor.  Nine from the set of AAS prizewinners are in the NAS.
	Focusing on this set of 16 who have won major honors from the AAS and are both in the top 50 in Tori25 and Read10 we find that 10 individuals have principally worked at a university and 6 have principally worked in a non-university setting.
	Of the 10 university based persons 8 are NAS members, while of the 6 non-university based persons only one is in the NAS.  Forming the 2x2 correspondence table, and evaluating the chi-squared statistic we find that this would occur by pure chance less than 1.4% of the time.  We thus establish with reasonable likelihood that elections of astronomers to the NAS are biased against non-university researchers; essentially the "affiliation bias" discussed by Lee et al (2013) for peer review of journal articles.  A similar analysis of the AAS prize group finds no bias.

**Discussion**
	Quantitative measures, AKA metrics, are increasingly used in evaluations of the past, present and potential scholarly performance of people and groups of people.  How well do these measures actually perform?  In this paper we have attempted to evaluate the performance of the principal scholarly metrics (citations (Tori), downloads (Read10), and peer evaluations) in measuring the past, present and potential performance of individual astronomers.
	While many measures have an air of exactitude, this author has 5246 citations, these papers had 5987 downloads, … , these numbers are meaningless outside their contexts.  Is a person with 5246 citations better (in some sense, and by how much) than a person with 3912 citations, or a person who had 5987 downloads last year?
	The basic performance questions for metrics are: given measurements on two individuals (A and B), can one assert that A is better than B in terms of past, present and/or (and especially) potential scholarly production or impact, and at what confidence level?  This is a different question than asking about the accuracy of the counts/measures themselves.  Here the question is, if the counts are perfectly accurate, how well can they represent the desired quantity: scholarly performance?
	There are several factors that strongly influence citation and download measures, but which are only weakly related to a person's current and future scholarly performance.  Age, scholarly discipline, and co-authorship practices vary widely among individuals, and, in practice, can only be imperfectly removed.
	Here we use statistics and a sample designed to minimize the effect of systematics, and reveal the underlying ability of citation and download statistics to measure/predict the past, present, and future scholarly performance of an individual.
	By comparing different measures on exactly the same sets of papers by exactly the same individuals in a sample designed to remove low performers or individuals with systematic differences we have shown the absolute limit on the ability of citation and download measures to measure individual differences to be 0.15-0.20 dex (1-$\sigma$).  To be able to say, at the 95% (2-$\sigma$) confidence level, that one person has had a greater impact than another the difference in their measures must be greater than 0.3 dex, a factor of two.

By using measures designed to be insensitive to changes in subfield (Tori) and number of collaborators (Tori and Read10) we have been able to show that changes in a person's research output with time are the dominant factor in comparing scholarly impact over different time periods. Different measures, such as citations (Tori) and downloads (Read10) weigh different time periods differently, and thus can get very different results for the same person.

As an example, even with the highly selected AA sample, the scatter about the mean relation between the contemporaneous impact measures Tori25 and Read10 is 0.347 (1-$\sigma$), this implies, e.g., that to assert with 95% confidence, that a person (A) who has a higher citation count than another person (B) will also have a larger download count, person A's citation (Tori) score must be (at least) a factor of five larger than person B's.

While the ability of these metrics in real situations to differentiate the scholarly impact of specific individuals with high confidence is obviously severely constrained, they do have statistical predictive power. Figure 16, for example, shows that individuals in the top 10% of Tori5 account for 46% of the individuals in the top 10% of Read10 more than 20 years later.

As a practical matter, personnel decisions are made amongst a group of candidates, not the entire set of persons with the relevant academic degree. We can model a group a candidates for top research positions by the very highly restricted subset of 49 persons who are both in the AA sample and in the top 10% of Tori5. The (1-$\sigma$) scatter about the mean relation of Tori5 and Read10, essentially the ability of Tori5 to predict behavior two decades later, is 0.395 dex, a multiplicative factor or 2.5; citation count differences between individuals smaller than this are simply not significant.

By examining measures of scholarly performance taken at different stages in the careers of a single age cohort in a single pure research discipline we have been able to compare and quantify the ability of the three primary means of scholarly evaluation (citations, downloads, and peer review) to measure current and predict future performance. We find each to be approximately equally effective, each to have a large variance, and each to achieve different results than the others.

A direct policy implication from these results is that although peer opinions are well correlated with metric measures (Norris and Oppenheim, 2003) they do not give the same results; thus suggestions, such as by Abramo, Cicero & D'Angelo, (2013) that peer evaluations be eliminated in favor of metrics only methods would lose significant information. We recommend that all three methods be used, taking properly into account the rather large scatter involved in each. Wilsdon, et al (2015) reach a similar conclusion.

Another, somewhat amusing result comes from the equivalence in effectiveness of citations with peer evaluations. The widespread use of the Impact Factor (Garfield 1972) of a journal to rank specific articles by an individual is now widely viewed as improper (e.g. Adler, Ewing, and Taylor 2009). As citation counts tend to follow a power law distribution, and with about 70% of all articles having citation counts below the mean, this would seem a reasonable view. However acceptance of an article by an established journal is not a citation measure, it is a peer evaluation measure.

Peer evaluations can certainly be ranked. Being offered a position at a top ranked research organization is different from being turned down, and is also different from receiving an offer from a lesser-ranked organization. Similarly having an article accepted by a highly selec-tive publication is different from having it rejected, and is different from having it accepted by a less selective publication. The Impact Factor effectively ranks the peer evaluation processes of the journals, and provides the numerical means to compare the result of the peer evaluations of individual articles by journals with citation studies.

**Conclusions**

Making personnel and funding decisions are likely the most important and most difficult aspects of scholarly life. These decisions are typically made by peer groups, now often informed by citation measures, and increasingly by other alternative metrics, principally article download based measures. How well can these metrics help employers and funders to find the "best" candidates?

By performing *a posteriori* corrections and selections to eliminate the effects of various biases and systematics, and by comparing individuals against themselves we have been able to determine the measurement uncertainty involved in comparing scholars using either citations or downloads. We find this uncertainty to be ~0.15 dex, where the underlying distribution is fully consistent with being log-normal. This means that, even if one were able to totally remove the effects of differing age, discipline, co-authorship, etc., to be able to state with 95% confidence that one person's citation or download counts are significantly larger than another's, those counts would have to be a factor of two larger.

By retrospectively comparing measures of specific individuals with measures of the same individuals with different measures at different times we have been able to quantify both the relative stability and the ability of the metrics to predict future performance.

We have found that the three methods, peer review, citation ranking, and download ranking are approximately equally able to predict future performance, but have quite different properties, and yield different results (in terms of the exact individuals chosen).

We have found that differences in metric scores between individuals that are less than a factor of 2.5 are not significant as predictors of future performance. As this result comes from a very strongly *a posteriori* selected sample we suggest as a rule of thumb that in real world comparisons a factor of three in citations or downloads be used as a threshold, less than that being an insignificant difference.

**Acknowledgements**

We have benefitted from conversations with Margaret Geller and Paul Ginsparg. Barbara Elwell validated the list of PhDs and their publications. We especially thank the ADS team, led by Alberto Accomazzi. The ADS is funded by NNX12AG54G. The work of the two referees has led to a substantial improvement of this paper.